\documentclass[useAMS,usenatbib]{mnras}
\usepackage{graphicx}
\usepackage{amsbsy,amsmath}
\usepackage{graphics,graphicx}
\usepackage{subfloat}
\usepackage[compatibility=false]{caption}
\usepackage{subcaption}
\usepackage[T1]{fontenc}
\usepackage{ae,aecompl}
\usepackage{newtxtext,newtxmath}
\usepackage[compact]{titlesec}

\title {A hot X-ray filament associated with A3017 galaxy cluster}  
\author[Parekh et al.] {Viral Parekh$^{1}\thanks{E-mail:
viral@rri.res.in}$,  F. Durret$^{2}$, P. Padmanabh$^{3}$, M. B. Pandge$^{4}\thanks {DST INSPIRE Faculty}$\\
$^{1}$Raman Research Institute, C. V. Raman Avenue, Sadashivnagar, Bangalore 560080, India.\\
$^{2}$ Institut d'Astrophysique de Paris, CNRS, 98bis Bd. Arago, 75014 Paris, France\\
$^{3}$Inter-University Centre for Astronomy and Astrophysics, Post Bag 4,
Ganeshkhind,  Pune 411007, India \\
$^{4}$Dayanand Science College, Barshi Road, Latur, Maharashtra 413512, India\\
}

\date{Accpted}
\pubyear{}
\begin{document}
\label{firstpage}
\maketitle

\begin{abstract}
\par Recent simulations and observations have shown large scale filaments in the cosmic web connecting nodes, with accreting materials (baryonic and dark matter) flowing through them. Current high sensitivity observations also show that the propagation of shocks through filaments can heat them up, and make filaments visible between two or more galaxy clusters or around massive clusters, based on optical and/or X-ray observations. We are reporting here the special case of the cluster A3017 associated with a hot filament. The temperature of the filament is 3.4$^{-0.77}_{+1.30}$ ~keV and its length is $\sim$ 1 Mpc. We have analysed its archival {\it Chandra} data and report various properties. We also analysed GMRT 235/610 MHz radio data. Radio observations have revealed symmetric two-sided lobes which fill cavities in the A3017 cluster core region, associated with central AGN. In the radio map, we also noticed a peculiar linear vertical radio structure in the X-ray filament region which might be associated with a cosmic filament shock. This radio structure could be a radio phoenix or old plasma where an old relativistic population is re-accelerated by shock propagation. Finally we put an upper limit on the radio luminosity of the filament region.    
\end{abstract}



\section{Introduction}

\par Recent cosmological n-body  simulations show that strong concentrations of  matter appear to be  interlinked by large, low density filaments \citep{2005Natur.435..629S,2008SSRv..134..311D}, surrounded by almost 
empty voids. This overall  pattern   is commonly known as the `Cosmic Web'. According to 
the $\Lambda$CDM  model  the  highest  mass concentrations reside  at   the  intersection  points of  dark matter filaments, through which they accrete both  dark and baryonic mass and  grow over time. {These interconnecting filaments have too  low a  density contrast  to be easily  detectable by  weak lensing \citep{2012Natur.487..202D}}. However, when  two  extremely  large mass concentrations, such as two galaxy clusters, come  relatively close to each other ($\sim$~order of Mpc distance), the  mass  density of the filament/bridge  between them is expected to be higher than the mean and might be detectable. Hence, binary  galaxy  clusters are the  ideal  locations to search for  large-scale  matter filaments and understand  their role in structure formation. Furthermore, galaxy clusters  grow in mass due to (i) merging with other cluster(s) and (ii) accretion of  matter. The first process leads to the  X-ray  substructures  due to  heating and mixing of intracluster  gas, and the later process  proceeds  via  supersonic  inflows near the virial radii of  clusters. This accretion process  happens predominantly through the filaments  which  are  connected to the clusters in their outskirts. Study of  filaments  is extremely important to understand cluster growth processes at the filamentary nodes and the acceleration of  high energy particles via shocks and turbulence. 

\par Current and future X-ray observations will allow us to select galaxy cluster targets to study filaments. High quality Chandra and XMM-Newton X-ray observations have already detected filaments these last years \citep{2003A&A...403L..29D, 2008A&A...482L..29W, 2013A&A...550A.134P}. Recently \cite{2015Natur.528..105E} have detected filamentary structures associated with the A2744 galaxy cluster with the Chandra X-ray telescope. These filaments are in agreement with the positions of over-densities of cluster galaxies and dark matter. This result supports evidence for the hypothesis
in which a large fraction of the missing baryons of the Universe reside in the filaments of the cosmic web. Previously, \cite{1999A&A...341...23K} have detected filamentary X-ray structure in the core region of the Shapley supercluster (between A3556 and A3558), \cite{2008A&A...482L..29W} have detected hot gas in the filament connecting the clusters of galaxies A222 and A223, and \cite{2009ApJ...695.1351B} have also reported filaments between A2804 and A2811 in the Sculptor supercluster based on X-ray absorption lines. 

\par X-ray observations with subarcsecond resolution also revealed a variety of substructures - cavities, bubbles, spiral features, cold fronts, etc. in the core of galaxy clusters \citep{2002AstL...28..495V,2007PhR...443....1M,2007ARA&A..45..117M,2011ApJ...737...99B, 2012NJPh...14e5023M,2013ApJ...773..114P}. The central active galactic nucleus (AGN) plays an active role in creating cavities and bubbles. In an investigation of a large sample of such systems, it was revealed that a central radio source was present in every system, with the radio plasma (or radio lobes) often filling the cavities and bubbles. The X-ray cavities and bubbles are therefore interpreted as regions where the X-ray emitting gas has been displaced by radio plasma produced by energetic outflows from the AGN that injects significant power into the ICM. Hence the central AGN is thought to be a heating source providing feedback to compensate for the excessive cooling of the cluster core. 
 
\par  Spiral features are likely related to ``sloshing'' of the ICM in the core of the cluster \citep{2011ApJ...737...99B}.  Simulations \citep{2006ApJ...650..102A, 2011ApJ...743...16Z} have shown that the off-centre collision of a cluster with a galaxy group, with mass ratios between 1/2 to 1/10, accelerates both the dark matter component and the cool gas of the cluster core, but the gas component is decelerated by ram pressure, resulting in a separation between the dark matter and the baryons. As the ram pressure weakens, the cool core gas falls back into the potential well, but overshoots it and begins to ``slosh''.   Heating of the ICM may also result from gas ``sloshing'' in the cluster central potential well. However, it has been noticed that this mechanism is efficient to move the cool gas from the centre but it cannot stop or regulate the cluster cooling-flow. 

\par A3017 ($z$ = 0.2195) is a peculiar cluster system  which we  discovered to be clearly  displaying  the  extremely rare occurrence  of  a large-scale ($\sim$ Mpc) inter-cluster  matter  filament.  The hot, gaseous  filament extending on Mpc-scale was discovered in {\it Chandra}  X-ray  observations (above 5$\sigma$ detection). This  filament is  much hotter and more X-ray luminous than the filamentary bridge between the A222/A223 cluster pair  \citep{2008A&A...482L..29W}. In addition to this, a pair of AGN related X-ray cavities  at  the  core region of A3017, one infalling group and several other groups interacting within A3017  are also detected. All these features make A3017 a prime target for  studies. We also obtained its 235/610 MHz dual frequency GMRT data. We have  listed the  properties  of A3017 in Table~\ref{sample}. We assume $H_{0}$ = 70 km s$^{-1}$ Mpc$^{-1}$ $\Omega_{M}$ = 0.3 and $\Omega_{\Lambda}$ = 0.7 throughout the paper. At redshift of 0.2195, the physical size of 1$''$ is  3.55 kpc.

\vspace{12pt}

\section{X-ray and Radio observations}

\subsection{X-ray data analysis}
\par We used {\it Chandra} archival data to study A3017 (Table \ref{xraydata}).  We followed standard X-ray data analysis and processed the data with CIAO 4.8 and CALDB 4.7.0. We first used the \verb"chandra_repro" task to reprocess all ACIS imaging data, removed high background flares (3$\sigma$ clipping) with the task \verb"lc_sigma_clip", and combined multiple data sets with \verb"merge_obs" script. All event files included the 0.3--7 keV broad energy band and 2$''$ pixel binning. We removed point sources around the cluster and filament region. We divided the count image by the exposure map and generated the flux image (photon~cm$^{-2}$~s$^{-1}$).  For the spectral analysis work, we used the CIAO Sherpa V1 package. For background estimation, we used {\it Chandra} blank sky files. 

\begin{table*}
\centering
\begin{small}
\caption{A3017 properties.}
\begin{tabular}{ccccccccccccccc}
\hline
\hline
Cluster & RA(J2000) & DEC(J2000) & $z$ & Luminosity & Y$_{X}$ $M_{500}$\\
 name   & h~m~~s & d~m~s & & $10^{44}$ erg~s$^{-1}$ (0.1--2.4 keV)& 10$^{14}$ $M_\odot$ \\
\hline
A3017  & 02 25 52 & -41 54.5 &0.2195 &10.17 & 5.81 \\
\hline
\end{tabular}
\label{sample}
\end{small}
\end{table*}

\begin{table}
\centering
\caption{A3017 {\it Chandra} X-ray data.}
\label{xraydata}
\begin{tabular}{ccccc}

\hline
           Obsid   &   instrument  &      obsdate       & exposure time (ks) \\
\hline
   15110&  ACIS-I & 2013-05-01   &             14.8 \\
   17476& ACIS-I & 2015-04-21   &              13.8 \\
\hline 
\end{tabular} 

\end{table}

\subsection{Radio data analysis}
\par We observed the A3017 cluster in dual frequency 610/235 MHz band with GMRT in August 2015 for a total duration of $\sim$ 6 hours (P.I. V. Parekh, obsid. 28$\_$087). This allows to record one polarization in each frequency band (610 MHz with RR and 235 MHz with LL). The GMRT software back-end with a bandwidth of 12 and 32 MHz at 235 and 610 MHz frequencies, respectively were used in these observations. The data were processed with a fully automated pipeline based on the SPAM package \citep{2009A&A...501.1185I,2014arXiv1402.4889I}, which includes direction-dependent calibration, modeling and imaging to suppress mainly ionospheric phase errors. In summary, the pipeline consists of two parts: a \emph{pre-processing} part that converts the raw data from individual observing sessions into pre-calibrated visibility data sets for all observed pointings, and a \emph{main pipeline} part that converts pre-calibrated visibility data per pointing into stokes I continuum images. The flux scale is set by calibration on 3C48 using the models from \citet{2012MNRAS.423L..30S}. We used default robust -1 weighting value in the SPAM to make images. More details on the processing pipeline and characteristics of the data products are given in paper on the first TGSS alternative data release \citep[ADR1]{2017A&A...598A..78I}.

\subsection{Images}

\par
\begin{figure*}
    \centering
    \begin{subfigure}[t]{0.48\textwidth}
        \includegraphics[width=\textwidth]{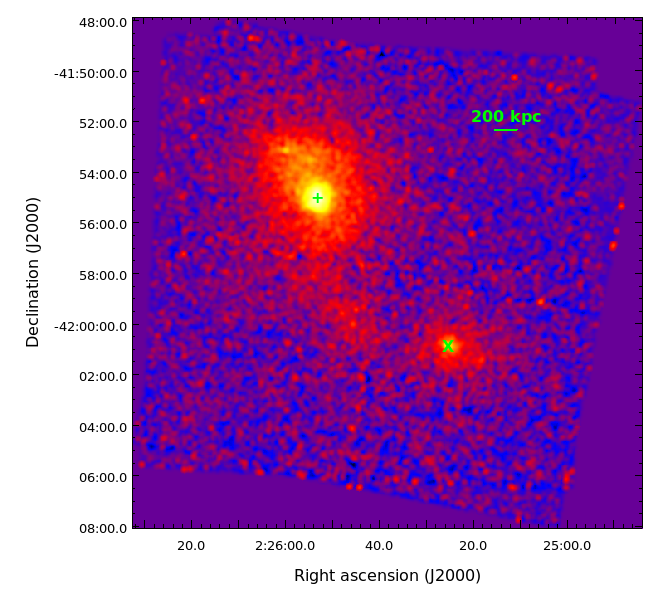}
        \caption{}
        \label{rfidtest_xaxis1}
    \end{subfigure}
    \begin{subfigure}[t]{0.51\textwidth}
        \includegraphics[width=\textwidth]{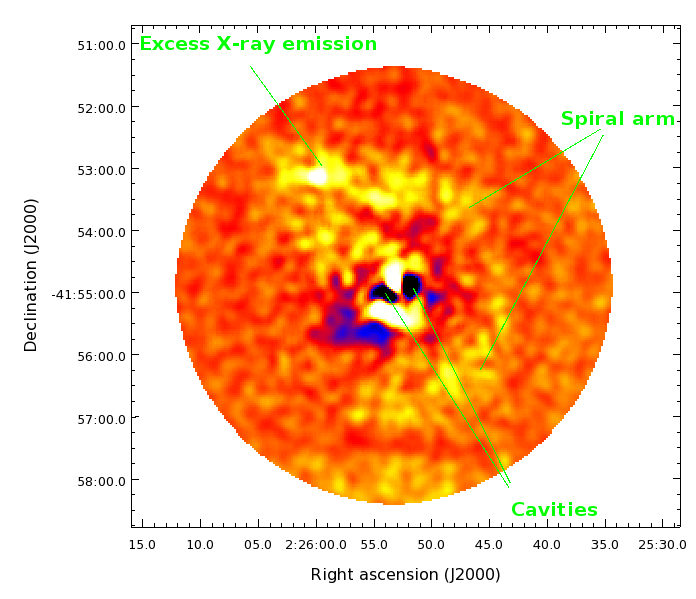}
        \caption{}
        \label{rfidtest_yaxis2}
    \end{subfigure}
     \begin{subfigure}[t]{0.48\textwidth}
        \includegraphics[width=\textwidth]{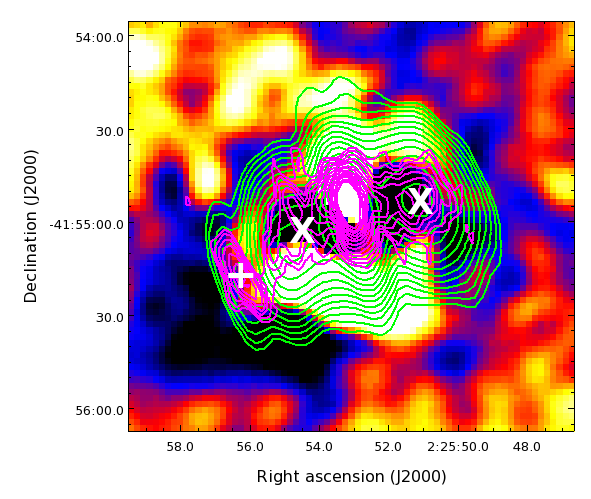}
        \caption{}
        \label{rfidtest_yaxis2}
     \end{subfigure}     
       \caption[]{(a) Background subtracted and exposure corrected {\it Chandra} ACIS-I X-ray image.  A3017 is marked with a `+' and the south X-ray cluster with a `x' sign. (b) Central core region of A3017. (c) Zoom of the A3017 core region with radio contours (green for 235 MHz and magenta for 610 MHz GMRT data). In this image, we marked the lobes (`$\times$') and unresolved radio source (`+').   The first radio contour is drawn at 5$\sigma$ and the levels increase in steps of $\sqrt{2}$. At 235 MHz, 1$\sigma$ rms is $\sim$ 0.45 mJy/beam and at 610 MHz 1$\sigma$ rms is $\sim$ 60 $\mu$Jy/beam. Both radio images are shown with their original resolutions. Resolution of 235 MHz image is 25$''$ $\times$ 9$''$, pa=14.5$^{\circ}$ and 610 MHz image is 10$''$ $\times$ 4$''$, pa=14$^{\circ}$.}
    \label{cluster_sample1}
\end{figure*}

  As shown in Fig.~\ref{cluster_sample1} (a), there is a filament of $\sim$ 1 Mpc projected size visible in X-rays, which seems to connect the A3017 (north) and south X-ray clusters. The core to core (X-ray peak to peak) distance between them is $\sim$ 470$''$, corresponding to a physical size of $\sim$ 1.7 Mpc. There is an Abell cluster, A3016 situated at R.A.: 02h25m22.1s and Dec: $-$42d00m30s. However, it is unclear that the south X-ray cluster is associated with A3016 or not. There is an offset of $\sim$ 45$''$ ($\sim 160$~kpc at the cluster redshift) between the X-ray peak of the south cluster and the position of A3016. See section~\ref{A3016_redshift} for the redshift analysis of A3016.  

 We used double 2D beta models to subtract the surface brightness model from the cluster X-ray image. Its residual  shows several interesting features. In addition to the filament between clusters, there are two symmetric cavities present in the core of A3017 (Fig.~\ref{cluster_sample1} (b)). Further there are two X-ray peaks lying in the north-south direction in the core of the cluster: a northern peak associated with radio and optical counterparts which could be a BCG, and a southern peak associated with X-ray emission. The distance between these two peaks is $\sim$ 12$''$.  Two X-ray cavities situated around the two peaks within $\sim$20$''$ from the BCG may indicate a  young AGN from which  the  jet and  lobe expansion has  just started. These cavities are very likely large empty bubbles evacuated by the AGN radio lobes.  In radio 235 and 610 MHz GMRT data (Fig.~\ref{cluster_sample1} (c)), we can see two radio lobes around the BCG and they fill the opposite X-ray cavities (marked with `$\times$'). We noticed that the east and west lobes have similar sizes, but the west lobe is more powerful than the east lobe. We also marked an unresolved radio source (left to the east lobe) in this radio image, with `+'. In Table~\ref{radio_prop} we listed the radio properties of the lobes and central radio source. We found steep spectra for the lobes and central radio source, with spectral indices of $\sim -2.6$, $\sim -3.3$ and $\sim -1.3$ for the west lobe, east lobe and central radio source, respectively.

\par  In this residual image, as shown in Fig.~\ref{cluster_sample1} (b), there is an excess X-ray emission visible in the north-east direction of A3017. It could be associated with infalling groups of galaxies, as suggested in the case of Abell~85 \citep{2005A&A...432..809D}. This excess X-ray emission connects with the cluster core region in the east-south direction, and towards the west connects with the large spiral arm extending towards the south around the cluster core. The radius of this spiral arm is $\sim$ 110$''$. This spiral arm could be connected with the inter-cluster filament at the cluster outskirts. Moreover, we  also noticed a  very  peculiar,  narrow  and extended  radio structure in the filamentary X-ray bridge region in  both  610 and 235 MHz radio images,  as shown in the zoom in the image subset of Fig.~\ref{radio_ver}. The angular (linear) size of this radio structure, placed  perpendicular to the filament axis, is $\sim$ 85$''$ (300 kpc) and $\sim$ 65$''$ (230 kpc) at 610 and 235 MHz, respectively. The calculated flux densities of the vertical radio structure are 5.0 $\pm$ 0.9 mJy at 235 MHz and 3.8 $\pm$ 0.5 mJy at 610 MHz. The integrated spectral index is $\sim -0.3$, which indicates a flat spectrum for  this structure and is significantly flat if this a typical extended radio galaxy.  We confirmed that there is no corresponding optical  counterpart  within 30$''$ of the strongest northern  peak (Fig.~\ref{radio_ver}(c)). Furthermore, at this position there is no radio source identified in the 843 MHz SUMSS catalogue or 150 MHz TGSS. 


\begin{table*}
\centering
\caption{A3017 radio lobes and BCG properties.}
\label{radio_prop}
\begin{tabular}{ccccccc}
\hline
Frequency & region  & size & flux density & radio power\\
\hline
MHz & & $''$ $\times$ $''$ & mJy & W Hz$^{-1}$ \\
\hline
\hline
235&            west lobe& 20$\times$10 &303 $\pm$30  & 25.63 \\
      &            east lobe&  24$\times$9 & 250 $\pm$ 25  &25.55\\
      &            BCG&       9$\times$5 & 514 $\pm$51       \\
610&            west lobe& 26$\times$16 & 25 $\pm$2,5   &24.55 \\
      &            east lobe&  19$\times$13 & 10 $\pm$1.0  &24.15\\
      &            BCG&   7$\times$5 &138 $\pm$ 13          \\
\hline
\end{tabular} 

\end{table*}


\begin{figure*}
    \centering
    \begin{subfigure}[t]{0.47\textwidth}
        \includegraphics[width=\textwidth]{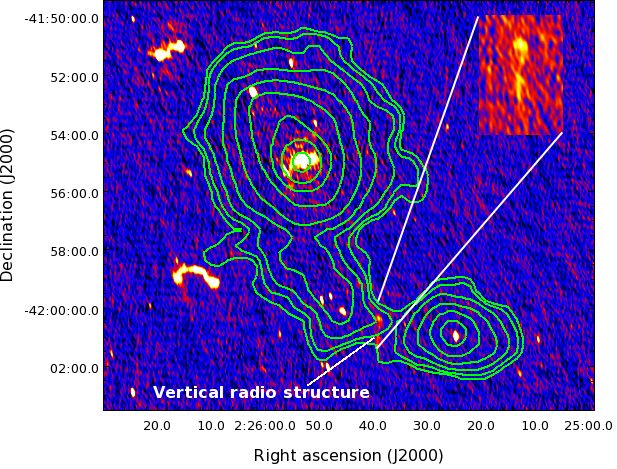}
        \caption{}
        \label{rfidtest_xaxis1}
    \end{subfigure}
    \begin{subfigure}[t]{0.48\textwidth}
        \includegraphics[width=\textwidth]{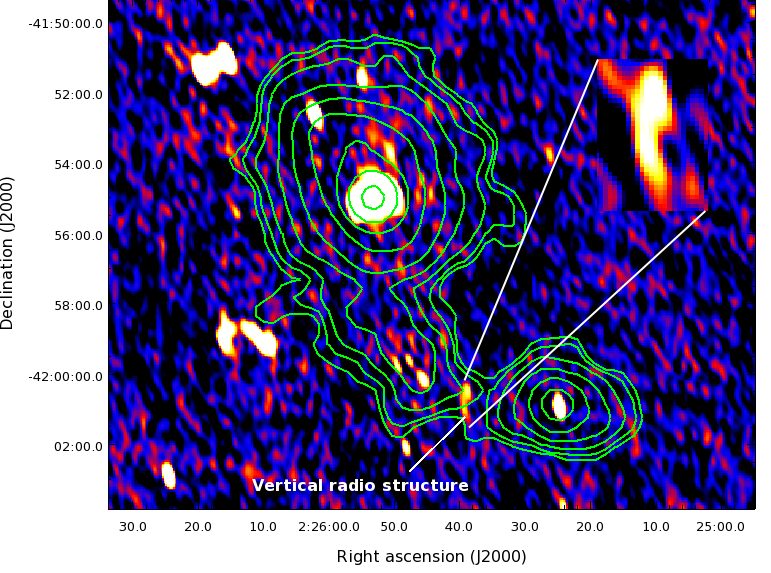}
        \caption{}
        \label{radio_ver}
    \end{subfigure}
       \begin{subfigure}[t]{0.45\textwidth}
        \includegraphics[width=\textwidth]{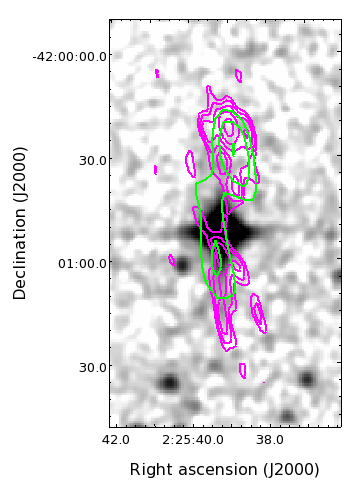}
        \caption{}
        \label{radio_ver}
    \end{subfigure}     
       \caption[]{(a) High resolution 610 MHz radio image (colour) and Chandra 
       X-ray contours (green). (b) High resolution 235 MHz radio image (colour) and  Chandra 
       X-ray contours (green). In  both images the peculiar  vertical radio structure 
        on the X-ray bridge region is highlighted within the inset. (c) Contours of the vertical radio structure at 235 (green) and 610 MHz (magenta) superimposed on the optical SuperCOSMOS image. There is no optical counterpart associated with this structure. The first radio contour is drawn at 3$\sigma$ and the levels increase in steps of $\sqrt{2}$. The rms noise values are the same as mentioned in the caption to Fig. 1.}
    \label{radio_ver}
\end{figure*}

\subsection{X-ray spectral analysis}
\par
\begin{figure*}
\centering
 \includegraphics[width=3in]{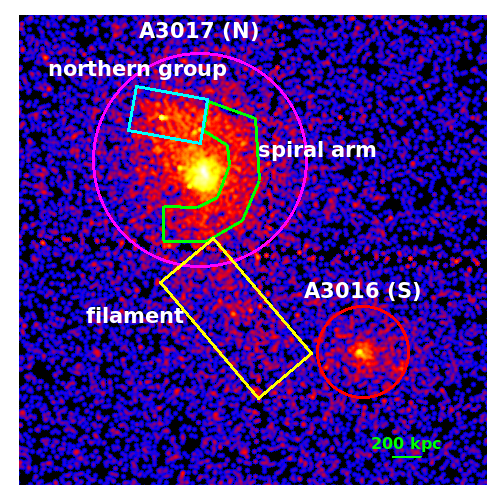}
\caption{A3017 X-ray image showing the different regions from which X-ray spectra were extracted.}
\label{spectra_region}        
\end{figure*}

  We extract the spectra from the A3017, filament, northern infalling group, spiral arm, and south X-ray cluster regions after subtracting the background (Fig.~\ref{spectra_region}). Since we do not have enough counts in our data, estimating the temperatures of the cavities is very difficult and leads to high uncertainties. { However, we estimated the significance for the detection of the filament around A3017.  We calculated that the total counts within the filament region (see below) are  $\sim$ 1750 and the average background counts are $\sim$ 1050, leading to S/N of $\sim$ 16}. We apply an apec model with photoelectric absorption (wabs) for which we have fixed the hydrogen column density to 0.02 $\times$ 10$^{22}$ atom cm$^{-2}$, the metal abundance to 0.3 solar, also reduced to 0.2 and 0.1 for the filament, and the redshift to 0.2195. {In this analysis, we used the solar abundance table in the CIAO Sherpa spectral fitting package from \cite{2009ARA&A..47..481A}}. We grouped spectral data so that each bin has at least 20 counts. In our spectral fitting we considered the energy range between 0.5 and 2.5 keV. We simultaneously fit the spectra from both  {\it Chandra} observations (Table \ref{xraydata}). 

\par To extract the spectra from the filament, we used a box of size ($l$ $\times$ $b$) 135$''$ $\times$ 300$''$ (480$\times$1065 kpc$^2$, $r$ = 68$''$). We found a temperature of 3.4$^{ -0.8}_{+1.3}$ for Z = 0.3 solar. The corresponding density is 7.7$^{-0.24}_{+0.27}$ $\times$10$^{-4}$ cm$^{-3}$.  Since we do not know the abundance fraction in the low density filament region, we tried two other values. For Z = 0.2 solar, the temperature is 3.1$^{-0.71}_ {+1.14}$ and the density is  7.8$^{-0.29}_{+0.32}$ $\times$10$^{-4}$ cm$^{-3}$. For Z = 0.1 solar, the temperature is 2.8$^{-0.63}_ {+1.12}$ and the density is 8.1$^{-0.38}_{+0.43}$ $\times$10$^{-4}$ cm$^{-3}$. We used circular regions of radii 3.5$'$ and 1.5$'$ for the A3017 (north) and south X-ray clusters, respectively to extract the spectra. For the northern group we took a 86$''$ $\times$ 142$''$ box region and for the spiral arm a 360$^{\circ}$ polygon region around the arm region, as shown in Fig.~\ref{spectra_region}. {We have listed the temperature, normalization and flux density values for the five regions in Table~\ref{xray_spectra} with 1$\sigma$ uncertainties.}

\begin{table*}
\centering
\caption{A3017 {\it Chandra} spectral analysis. These values are measured in soft energy (0.5-2.5 keV) band.}
\label{xray_spectra}
\begin{tabular}{ccccccc}
\hline
Cluster region   & counts & $\chi^{2}$/dof & temp & normalization & flux \\
\hline
& &  &keV& 10$^{-4}$&10$^{-13}$ cgs \\
\hline
A3017 & 5611 & 171/241 & 5.8$^{-0.6}_{+0.6}$ &56$^{-0.7}_{+0.7}$  &32 \\
filament & 427 & 36/47 & 3.4$^{-0.8}_{+1.3}$  &  3.8$^{-0.2}_{+0.3}$  &2.1   \\
south X-ray cluster & 574 &40/54 & 3.7$^{-0.7}_{+1.2}$   &  5.1$^{-0.3}_{+0.3}$ & 2.8    \\
spiral arm & 1000 & 101/178 & 6.0$^{-0.8}_{+0.9}$ & 7.1$^{-0.3}_{+0.3}$ & 6.9 \\
northern group & 566 &90/104 & 3.8$^{-0.7}_{+1.4}$ & 6.4$^{-0.3}_{+0.3}$ &3.6 \\
\hline 
\end{tabular} 

\end{table*}

\section{Optical analysis}
\par This section investigates the optical properties of the galaxy cluster A3017 as well as the filament structure attached to it. This has been done by using available catalogues extracted from the WISE and ESO Red public surveys.
\par
\begin{figure*}
    \centering
    \begin{subfigure}[t]{0.46\textwidth}
        \includegraphics[width=\textwidth]{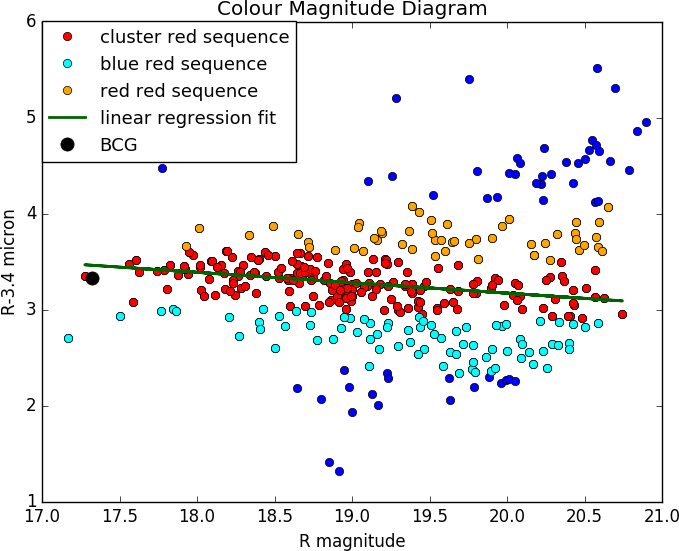}
        \caption{}
        \label{rfidtest_xaxis1}
    \end{subfigure}
    \begin{subfigure}[t]{0.48\textwidth}
        \includegraphics[width=\textwidth]{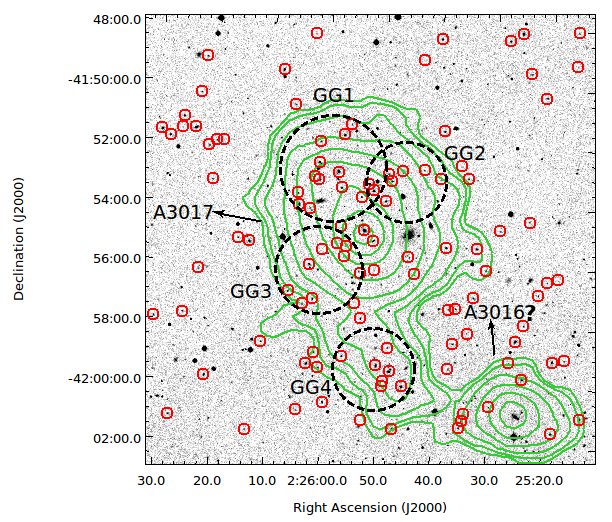}
        \caption{}
        \label{rfidtest_yaxis2}
        \end{subfigure}

    \caption{Left: R-3.4 $\mu$m colour against R magnitude for the galaxies in the A3017 area, taken from the SuperCOSMOS and ESO Red catalogues. The red filled circles show the galaxies belonging to the cluster red sequence, indicated by the green line, the yellow filled circles correspond to the ``red red sequence'' and the cyan filled circles correspond to the ``blue red sequence'' (see text). Right: SuperCOSMOS optical image  of A3017. Green contours show the X-ray emission and red circles show the member galaxies. {Black dashed circles} indicate possible groups of galaxies at the redshift of A3017.}
    \label{optical_img}        
\end{figure*}

 Galaxy clusters are well characterized by old and early-type members that correspond to red galaxies. Multiband optical data has been used to obtain the colour magnitude diagram (CMD) (Figure~\ref{optical_img}). For this, the WISE and SuperCOSMOS catalogues (ESO Red survey) were cross matched (tolerance radius = 6$"$). The R-3.4$\mu$m colour was plotted against the R magnitude. Only galaxies (corresponding to Class=1 in SuperCOSMOS catalogue) have been considered. Further, within the Class~1 sources, galaxies within the magnitude range R=17-21.5 were chosen; these are approximately the limits for the BCG magnitude and for the ESO RED-R survey magnitude limit. The most likely red sequence region in the CMD is in the range of colours ~3-3.5. The linear regression fit was applied to galaxies only within this magnitude range. A slightly negative slope (m = $-0.0574$) is found, indicating the variation of metallicity across the galaxies with magnitude.  We identify  groups of galaxies in the northern area of the A3017 cluster (black dashed circles) as well as the filament protruding from it. The majority of the galaxies in the cluster and above mentioned regions fall within the selected colour range.

\section{Do A3017 and A3016 form a pair of clusters?}
\label{A3016_redshift}

\begin{figure*}
    \centering
    \begin{subfigure}[t]{0.32\textwidth}
        \includegraphics[width=\textwidth]{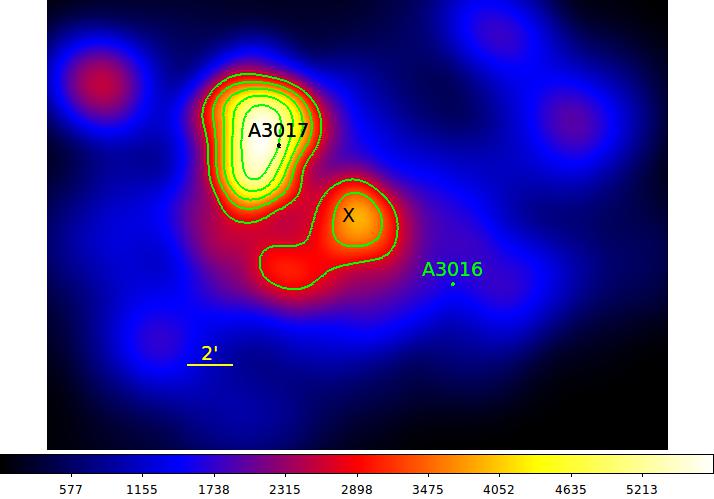}
        \caption{}
        \label{dens}
    \end{subfigure}
    \begin{subfigure}[t]{0.32\textwidth}
        \includegraphics[width=\textwidth]{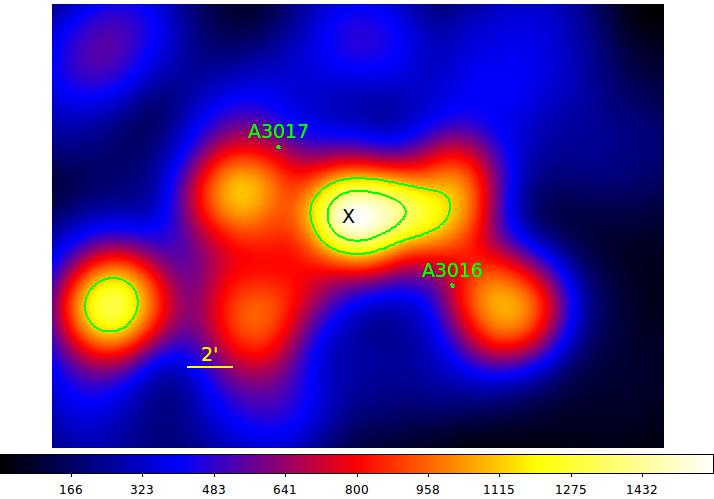}
        \caption{}
        \label{densblue}
        \end{subfigure}
    \begin{subfigure}[t]{0.32\textwidth}
        \includegraphics[width=\textwidth]{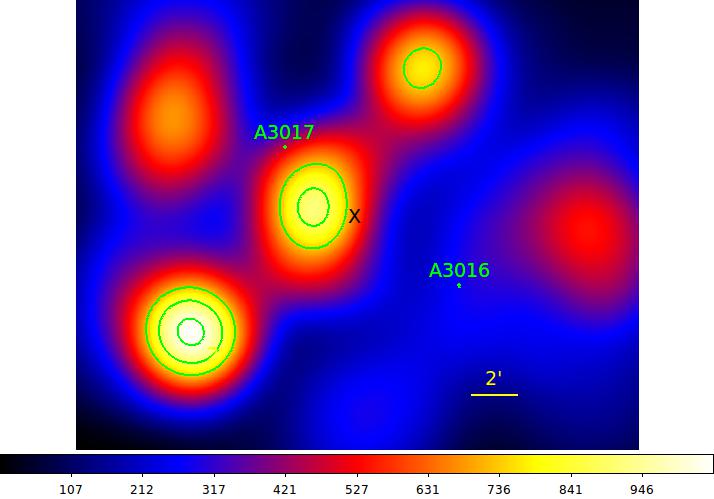}
        \caption{}
        \label{densred}
        \end{subfigure}

        \caption{Left: density map for the galaxies located along the
          red sequence of A3017 (the red points in
          Fig.~\ref{rfidtest_xaxis1}). Middle: density map for the
          galaxies located below the red sequence of A3017. Right:
          density map for the galaxies located above the red sequence
          of A3017. The significance contours start at 3$\sigma$ and
          increase by steps of $1\sigma$. The positions of A3017 and
          A3016 are shown with tiny circles. The position of
          G256.55-65.69 (see text) is indicated with an X. North is
          top and east is left. }
    \label{density}        
\end{figure*}

\par In order to see how the galaxies are distributed in the field, we
built a density map based on the red sequence galaxies of A3017. The
density map was computed with an adaptive kernel as explained in
Durret et al. (2016), with a pixel size of 18$"$ and 100
bootstraps, and the significance contours were computed as explained
in that paper. The corresponding density map is shown in
Fig.~\ref{density} (left).  We see that A3017 is detected at a
7$\sigma$ level. A second structure half way between A3017 and A3016
is also detected at a $4\sigma$ level, while we see absolutely no
signal at the position of A3016 (at the position given by NED).

This very puzzling result led us to question the redshift of A3016.
{First, we can note that Simbad gives a redshift $z$ = 0.2195 for
  this cluster \citep{2011A&A...536A...8P} }while NED gives no
redshift for this cluster. Second, there are extremely few references
on A3016 in the literature.

So we decided to extract from our catalogue the galaxies along a
``blue red sequence'' and a ``red red sequence'', with the same slope
and width of $\pm$ 0.3 as the A3017 RS, but located at colours 0.6~mag
below and above the RS respectively. The corresponding density maps
are shown in Fig.~\ref{density} (middle and right). We can see that
A3016 is detected in neither of them. The image built with galaxies
bluer than A3017 shows a quite intense structure between A3017 and
A3016. Less than 0.5$'$ from the peak of this structure is located a
galaxy cluster (hereafter cluster~X) named [BM78]~231 in NED (at
coordinates R.A.:02h25m40.1s, Dec:$-41$d57m31s, no redshift available). The
position of this cluster is marked with an X on the figure.  The image
built with galaxies redder than A3017 shows two intense structures,
one about 1.8$'$ east of cluster~X, and one further to the
southeast. We could not identify either of these structures with a
galaxy cluster in NED.

The two possible explanations to the fact that A3016 is not detected
in these density maps are: 1)~it is not a cluster, and the X-ray
detection as an extended source is wrong (which is rather surprising
in view of the good spatial resolution of Chandra); 2)~it is too
distant to be detected in our rather shallow optical data. In any
case, the hypothesis that we are observing a pair of clusters is most
probably not correct, and A3017 does not form a pair with A3016.

\section{Discussion and Conclusion}

\par There is a hot X-ray filament associated with A3017 cluster,
implying a significant baryon over-density in the filament
region. This over-dense filament appears to be connected with another
X-ray cluster.  We estimate temperatures of $\sim$ 5.8$^{-0.57}_{+0.62}$ keV
for A3017 and 3.7$^{-0.73}_{+1.22}$ keV for the south X-ray
cluster. {The temperature of group sized systems is usually $\lesssim$ 1 keV} \citep{1996astro.ph.11148B, 2002ARA&A..40..539R}, hence this X-ray emission is most
probably related to the cluster ICM than to a group. Current
multi-wavelength observations show that cluster mergers have several
stages \citep[and references therein]{2015A&A...575A.127P}; beginning with the early stage
when two bodies are approaching each other, followed by first
interaction, mixing of the hot gases, violent collision and shock
production, turbulence injection, ram pressure stripping, the
intermediate stage when the two bodies are receding from each other,
and finally the merger and relaxation on a time scale of a few
Gigayears.  From the optical data for A3017, it becomes clear that
A3017 is an early stage merger because the positions of the brightest
central galaxies are still associated with the X-ray peaks. In a late
stage merger, the galaxies and ICM are often distributed
differently. There could be a weak shock present in the A3017 filament
as in the cases of A222/A223 and CIZA J1358.9-4750. The presence of a
weak shock in the filament could be responsible for the linear
vertical radio structure visible in the radio map
(Fig.~\ref{radio_ver}). However, present X-ray data are not sufficient
to estimate the properties of the shock in the filament. In
simulations \cite{2010PASJ...62..335A} have shown that in the early
stage of a merger (t $\approx$ 0.5 Gyr), two clusters interact with
each other at the outskirts and then create linked regions (filaments
or bridges) where the ICM is compressed due to the collision. Further,
these authors have postulated that there exist shock fronts with Mach
numbers $M \sim 1.5 - 2.0$ at the positions of the bridge
regions. These shocks are immediately formed at the contact interface
of the two clusters when the clusters begin to collide, and they then
propagate in a direction perpendicular to the merger axis outside of
each cluster, forming outgoing merger shocks
\citep{2006Sci...314..791B}. However, in the present case, there is no
strong evidence that the south X-ray cluster is at the same redshift
as A3017, based on the galaxy density map analysis, so these two
clusters do not form a binary cluster merger. We need further optical
spectroscopic information to prove or disprove whether the X-ray
bridge is connecting two binary clusters or is only associated with
A3017 as a hot X-ray tail \citep{2004ApJ...611..164D,
  2006ApJ...637L..81S, 2015A&A...583L...2S}.


\par We estimate temperatures of 3.4$^{-0.77}_{+1.30}$  for the A3017
filament, assuming Z = 0.3 solar. The temperature of the filament is
hotter than expected, so the hot gas in the filament is probably
the ICM from the cluster outskirts, with some
contamination by cluster gas also. We calculated that the density of
the filament is between $\sim$ 7.5 $\times$ 10$^{-4}$ and $\sim$ 8.1 $\times$ 10$^{-4}$
cm$^{-3}$, its mass $\sim$ 4.2 - 4.5 $\times$ 10$^{12}$ $M{_\odot}$ and its
entropy $\sim$  320 - 405 keV cm$^{2}$, for metallicities between $Z_{\odot}$ = 0.3 and
0.1. To calculate these values, we assumed the volume of the filament
to be a cylinder of size 135$''$ $\times$ 300$''$ (480 kpc $\times$
1065 kpc).  Accretion shocks heat the cluster outskirts {($>$
$R_{200}$) \citep{2009ApJ...696.1640M}} raising entropies above 1000 keV cm$^{2}$. This entropy
is expected to increase further out to the virial radius. In this
case, the low entropy of the filament suggests that the radiating hot
gas in the A3017 filament has not yet been reached by the
shock-heating operating on the ICM in the cluster outskirts.

\par The core region of A3017 is cool, with an estimated central
cooling time of $\sim$ 1.4 Gyr\footnote{$t_{cool}$ = 69
  $\left(\dfrac{n_{e}}{10^{-3} cm^{-3}}\right)^{-1}$
  $\left(\dfrac{T}{10^{8} K}\right)^{1/2}$ Gyr}, and no sign of
disturbances in the X-ray surface brightness distribution. There are
substructures, two opposite cavities, and a spiral arm, visible in the
core region and surrounding it. The cavities are filled with radio
emission at 235 and 610 MHz, and they are very likely large empty
bubbles evacuated by the AGN's radio lobes. Current low resolution
radio data are not sufficient to resolve the jet emission from the
AGN. There is a spiral feature also visible which is in good agreement
with gas sloshing in connection with a sub-cluster or group merger.
There is also an excess of X-ray emission visible in the north-east
direction which is associated with subgroup GG1 as marked in Figure
\ref{optical_img}. We estimated the temperature of this group to be
3.8$^{-0.7}_{+1.4}$ keV, {which is more typical of a small
  cluster}.  This {small cluster} could be infalling on to the
main cluster body of A3017. The entropy of this infalling group is
$\sim$ 220 keV cm$^{2}$.  There is another subgroup GG2 also
associated with the large spiral arm.  Generally steep spectrum
mini-radio halos of the size $\sim$ 100-500 kpc are expected in cool
core clusters and believed to be directly connected with turbulent
re-acceleration in gas sloshing activity
\citep{2011MmSAI..82..632Z}. However, we have not detected any diffuse
radio emission in the present radio data.

\par The radio detection of filaments can probe the distribution of
cosmic rays and magnetic fields in the cosmic web
\citep{2017arXiv170205069V}, but such a detection is very challenging,
due to the extremely low magnetic field (10-100 nG) and to the
inefficiency of relativistic particle acceleration
\citep{2015aska.confE..97V}. However, simulations
\citep{2015A&A...580A.119V, 2012MNRAS.423.2325A} predict that it is
possible to detect filaments in radio observations if the shocks
accelerate the electrons sufficiently. In current observations, there
is no detection of radio emission in the filament region of
A3017. Hence we calculate the upper limit on the radio luminosity of
the filament. We calculated this upper limit (with the 610 MHz radio
data) by first measuring the rms in the X-ray filament assuming the
same cylindrical region as mentioned above. This rms is measured for a
single beam, so we need to scale it with a factor of
$\sqrt{(size^{2}/beam~area)}$ to take into account the number of beams
within the total filament size. Assuming the 135$''$ $\times$ 300$''$
filament size from the X-ray measurement, we found an upper limit to
the radio power of the filament of $<$ 23.95 W Hz$^{-1}$ at 610 MHz at
the 3$\sigma$ level.
  It will be important to observe such
over-dense and hot X-ray filaments around clusters and between cluster
pairs with future high sensitive radio telescopes such as SKA, MeerKAT
and ASKAP to constrain the magnetic field in filaments.


\section*{Acknowledgments}
We thank the referee for some useful comments.
MBP gratefully acknowledges the support of Department of Science and
Technology (DST), New Delhi under the INSPIRE faculty Scheme
(sanctioned No. DST/INSPIRE/04/2015/000108). This research has made use
of the data from {\it Chandra} Archive and of software provided by the
Chandra X-ray Centre (CXC) in the application packages CIAO, ChIPS,
and Sherpa.  This research has also made use of NASA's Astrophysics
Data System, of the NASA/IPAC Extragalactic Database (NED) which is
operated by the Jet Propulsion Laboratory, California Institute of
Technology, under contract with the National Aeronautics and Space
Administration, and of the SIMBAD database, operated at CDS,
Strasbourg, France.  Facilities: Chandra (ACIS), HST (ACS). We thank
the staff of the GMRT who have made these observations possible. GMRT
is run by the National Centre for Radio Astrophysics of the Tata
Institute of Fundamental Research. We also thank Huib Intema for
providing SPAM package for reducing GMRT data.

\bibliography{references}

\end{document}